\documentclass[letterpaper,prd,twocolumn,showpacs,showkeys,preprintnumbers,superscriptaddress,nofootinbib]{revtex4}

\usepackage{graphicx,color}
\usepackage{amsmath,amssymb,amsfonts}
\usepackage{stackrel}
\usepackage{tabularx}
\usepackage{url}
\newcolumntype{Y}{>{\centering\arraybackslash}X}
\usepackage{booktabs}
\AtBeginDocument{
    \heavyrulewidth=.08em
        \lightrulewidth=.05em
        \cmidrulewidth=.03em
        \belowrulesep=.65ex
        \belowbottomsep=0pt
        \aboverulesep=.4ex
        \abovetopsep=0pt
        \cmidrulesep=\doublerulesep
        \cmidrulekern=.5em
        \defaultaddspace=.5em
}
\usepackage{xcolor}
\usepackage{hyperref}
\hypersetup{
    colorlinks=true,
        linkcolor={blue!80!black},
        citecolor={blue!80!black},
        urlcolor={blue!80!black} 
}

\newcommand{\be}{\begin{equation}} \newcommand{\ee}{\end{equation}}
\newcommand{\bea}{\begin{eqnarray}} \newcommand{\eea}{\end{eqnarray}}

\begin{document}


\title{Finite-temperature Equations of State for Neutron Star Mergers}

\author{Paul~M.~Chesler}
\email{pchesler@g.harvard.edu}
\affiliation{Black Hole Initiative, Harvard University\\
Cambridge, MA 02138, USA}

\author{Niko~Jokela}
\email{niko.jokela@helsinki.fi}
\affiliation{Department of Physics and Helsinki Institute of Physics\\
P.O.~Box 64, FI-00014 University of Helsinki, Finland}

\author{Abraham~Loeb}
\email{aloeb@cfa.harvard.edu}
\affiliation{Black Hole Initiative, Harvard University\\
Cambridge, MA 02138, USA}

\author{Aleksi~Vuorinen}
\email{aleksi.vuorinen@helsinki.fi}
\affiliation{Department of Physics and Helsinki Institute of Physics\\
P.O.~Box 64, FI-00014 University of Helsinki, Finland}

\begin{abstract}

The detection of gravitational waves from a neutron star merger has opened up the possibility of detecting the presence or creation of deconfined quark matter using the gravitational wave signal. To investigate this possibility, we construct a family of neutron star matter equations of state at nonzero density and temperature by combining state-of-the-art nuclear matter equations of state with holographic equations of state for strongly interacting quark matter. The emerging picture consistently points toward a strong first order deconfinement transition, with a temperature-dependent critical density and latent heat that we quantitatively examine. Recent neutron star mass measurements are further used to discriminate between the different equations of state obtained, leaving a tightly constrained family of preferred equations of state.

\end{abstract}

\preprint{HIP-2019-17/TH}


\maketitle

\section{Introduction}

The simultaneous measurement of gravitational wave (GW) and electromagnetic (EM) signals from the apparent merger of two neutron stars (NSs) in August 2017 marked the birth of a new era of multimessenger astronomy \cite{TheLIGOScientific:2017qsa}. At the same time, this event also solidified the role of NSs as a laboratory for dense QCD, as demonstrated by the multitude of subsequent studies using this observation to constrain the properties of nuclear and even quark matter \cite{Margalit:2017dij,Rezzolla:2017aly,Annala:2017llu,Most:2018hfd,Zhou:2017pha,Paschalidis:2017qmb,Krastev:2018nwr,Zhu:2018ona,Annala:2017tqz,Tews:2018chv,Abbott:2018exr,Sieniawska:2018pev,Jokela:2018ers,Nandi:2018ami,Li:2018ayl,Tsang:2018kqj,Annala:2019puf,Baym:2019iky,Bai:2019jtl,Lim:2019som,Fasano:2019zwm,Zhou:2019omw,Oter:2019kig}.  The inspiral gravitational wave signal is consistent with fairly small tidal deformabilities and neutron star radii \cite{Abbott:2018exr}, indicating a relatively soft equation of state (EoS) at zero temperature.

What was not recorded in the GW170817 event was any trace of postmerger dynamics, which is likely due to the high frequencies involved in this part of the GW signal. Numerical simulations of mergers yield differentially rotating remnants with temperatures typically reaching tens of MeV (for reviews of this topic, see \cite{Clark:2015zxa,Baiotti:2016qnr,East:2019lbk}).
The vibrational modes of the remnant, which are sensitive to its size and the speed of sound, excite gravitational waves \cite{Takami:2014tva,Clark:2015zxa,East:2019lbk,Baiotti:2016qnr,Bauswein:2015yca,Bauswein:2015vxa,Stergioulas:2011gd}. Hence, information about the finite temperature  EoS is contained in the post-merger gravitational wave signal. A particularly interesting scenario, studied e.g.~in \cite{Most:2018eaw,Bauswein:2018bma}, is one where the two colliding stars initially contain no deconfined matter, but the system traverses through a phase transition region during the merger (note, however, that the authors of \cite{Bauswein:2018bma} in addition looked into mergers of hybrid stars). An obvious question then becomes, whether the phase transition and the appearance of deconfined matter may be observable in the post-merger gravitational wave signal \cite{Ibanez:2018myp,Alford:2019oge}. The results of \cite{Most:2018eaw,Bauswein:2018bma}, obtained using various model EoSs for nuclear and quark matter, indicate that the answer is yes. In order to confirm the robustness of these conclusions in future simulations, it is, however, very important to complement the EoSs used with further modern microscopic descriptions of both phases.

Optimally, the EoS of dense QCD matter should be determined using a single non-perturbative method, such as lattice QCD, which has indeed provided accurate results in the limit of high temperatures and small or vanishing chemical potentials \cite{Borsanyi:2013bia,Bazavov:2014pvz}. 
At finite density, lattice QCD unfortunately suffers from the so-called sign problem (see e.g.~\cite{deForcrand:2010ys} for a discussion of this topic), which ultimately means that this method is at present not suitable for generating an EoS for NS mergers. A robust but less accurate alternative is to generate families of NS matter EoSs by interpolating between reliable first principles calculations at low \cite{Tews:2012fj,Gandolfi:2009fj} and high \cite{Kurkela:2009gj,Gorda:2018gpy} densities. Such an approach has indeed been successfully followed at exactly zero temperature \cite{Hebeler:2013nza,Kurkela:2014vha,Annala:2017llu,Most:2018hfd}, with results that are becoming sensitive to the characteristics of the deconfinement transition in the $T=0$ limit \cite{Annala:2019puf}. At nonzero temperatures, these types of studies do not exist yet, which is largely due to the first ab initio nuclear theory study having appeared only very recently \cite{Carbone:2019pkr}, although its high-density counterparts have been available for some time \cite{Vuorinen:2003fs,Kurkela:2016was}. Recalling that the accuracy of the interpolation studies will in any case be fairly limited in the most interesting density interval, containing the deconfinement transition, it is clear that fundamentally new approaches to the physics of dense QCD matter are direly needed.

A promising nonperturbative tool to tackle the finite-density thermodynamics of strongly interacting matter  is the holographic duality \cite{Maldacena:1997re}. Its utility stems from the fact that it maps challenging strongly coupled quantum field theory problems onto classical partial differential equations in higher dimensions, which can be solved numerically.  Holographic models have been widely employed to study the dynamics of strongly coupled quark-gluon plasma produced at RHIC and the LHC; for a review, see \cite{CasalderreySolana:2011us}. Highlights include studies of e.g.~thermalization \cite{Chesler:2008hg,Chesler:2010bi,Chesler:2015fpa}, jet quenching \cite{Herzog:2006gh,Gubser:2006bz,CasalderreySolana:2006rq}, and transport coefficients  \cite{Kovtun:2004de,Bhattacharyya:2008jc,Baier:2007ix}.  Applications of holography to heavy-ion collisions have focused on the limit where the chemical potential is small compared to the temperature, and consequently few studies have been performed in the context of dense and cold QCD matter (for some exceptions, see however \cite{Hoyos:2016zke,Hoyos:2016cob,Ecker:2017fyh,Annala:2017tqz,Li:2015uea,Faedo:2017aoe,Faedo:2018fjw,Jokela:2018ers,Ishii:2019gta}). For this reason, we focus on the simplest nontrivial models of high-density QCD matter, namely bottom-up holographic models of the deconfined phase.

Perhaps the most refined holographic bottom-up model of QCD is the Veneziano limit ($N_f\sim N_c\to\infty$) of the improved holographic QCD (V-QCD) framework \cite{Gursoy:2007cb,Gursoy:2009jd,Jarvinen:2011qe,Alho:2013hsa}. This model has been designed to not only respect the correct symmetries of QCD and match the lattice QCD thermodynamics in the zero-density limit but also to have the right UV properties, including the perturbative running of the gauge coupling. Very recently, this setup has been analyzed in detail in the limit of high densities and small temperatures \cite{Jokela:2018ers}, which offers a way to non-perturbatively model the quark matter phase in NSs and their mergers. In this paper, our goal is to match the predictions of this model with state-of-the art EoSs for nuclear matter \cite{Typel:2009sy,Fattoyev:2010mx,Steiner:2012rk}, ending up with a family of EoSs for NS matter at nonzero temperatures. We have  chosen the low-density models to ensure both compatibility with all existing robust NS observations and availability of the resulting matched EoSs for a wide range of temperatures.

Our paper is organized as follows. In Section~\ref{sec:model}, we review our setup in some detail, concentrating in particular on the novel description of the quark matter phase via V-QCD. In Section~\ref{sec:results}, we then present our results for the matched EoSs, and analyze the resulting phase diagrams and the properties of the deconfinement transition. Finally, Section~\ref{sec:discussion} contains our conclusions, and in the ancillary material of the arxiv entry (v1) we provide our quark matter EoSs in a tabular format.

\section{Models and matching setup}\label{sec:model}

As explained above, we model the confining phase of QCD using state-of-the-art phenomenological EoSs that are tabulated for a sufficient set of nonzero temperatures. Our criteria for choosing the EoSs are that they should be maximally different, yet consistent with robust observational information concerning their $T=0$ limit (assuming no phase transition to quark matter): 
\begin{itemize}
 \item The $T=0$ EoSs must support the heaviest known NSs with $M\approx 2M_\odot$ \cite{Demorest:2010bx,Antoniadis:2013pzd},
 \item The $T=0$ EoSs  must produce tidal deformabilities for 1.4$M_\odot$ NSs consistent with the 90\% confidence limits provided by LIGO and Virgo  \cite{Abbott:2018exr}. Note that this constraint has been translated to limits for NS radii  \cite{Annala:2017tqz} with results that are in agreement with recent direct radius measurements; cf.~e.g.~\cite{Nattila:2017wtj}.
\end{itemize}
Of the EoSs available in \cite{Hempelpage}, these constraints are satisfied by the ``DD2'' EoS of \cite{Typel:2009sy}, the ``IUF'' EoS of \cite{Fattoyev:2010mx} (see also \cite{Fattoyev:2017jql}),  as well as the ``SFHx'' EoS of \cite{Steiner:2012rk}, of which the second is in mild tension with the observation of two-solar-mass stars. For simplicity of presentation, we implement local charge neutrality and beta equilibrium. We have, however, explicitly checked that the qualitative aspects of our results are insensitive to the electron fraction $Y_e$, and in addition note that we provide our quark matter EoSs for different fixed values of $Y_e$ in the ancillary material of the arxiv-version (v1) of this paper. Finally, we remark that in order to reproduce the proper mass-radius (MR) relations in these models at zero temperature, we use the crustal EoS of \cite{Negele:1972zp} at the lowest densities; for the majority of our discussion, this detail is, however, unimportant.

On the quark matter side, we employ the holographic model V-QCD, which can be viewed as a merger of two ingredients. Gluon dynamics is modeled through the five-dimensional improved holographic QCD model of \cite{Gursoy:2007cb,Gursoy:2009jd}, while the fundamental flavor degrees of freedom are added by the introduction of $N_f$ copies of the tachyonic DBI action \cite{Bigazzi:2005md,Casero:2007ae,Bergman:2007pm,Dhar:2007bz,Dhar:2008um,Jokela:2009tk,Alho:2013hsa}. We consider the quarks dynamical, i.e.~work in the Veneziano limit of the theory, keeping $N_c/N_f=1$ fixed in the limits $N_c\to\infty$, $N_f\to\infty$ \cite{Jarvinen:2011qe}. As usual, the 't Hooft coupling $g^2N_c$ is also kept fixed in this limit.  On the gravity side, this means that we need to consider the full backreaction of the flavor action.

Our gravitational theory is discussed in detail in \cite{Alho:2013hsa,Jokela:2018ers}, so here we merely outline the salient details. The gravitational theory consists of the metric $g_{\mu \nu}$, a $U(1)$ gauge field $A_\mu$,
a dilaton  $\lambda$ and a tachyon $\tau$.  The action reads
\bea
 S & = & N_c^2 M_{\rm pl}^3\int d^5 x\sqrt{-\det g}\left[R -\frac{4}{3}\frac{(\partial \lambda)^2}{\lambda^2}+V_g(\lambda)\right] \nonumber\\
  &&\!\!\!\!\!\!\!\!\!\!\!\!-N_f N_c M_{\rm pl}^3\int d^5 x V_{f0}(\lambda) e^{-\tau^2} \nonumber \\
   && \ \ \ \times \sqrt{-\det(g_{\mu\nu}+\kappa(\lambda)\partial_\mu\tau\partial_\nu\tau+w(\lambda)F_{\mu\nu})}\ ,\label{eq:SVQCD}
\eea
where $R$ is the Ricci scalar, $F_{\mu \nu}$ is the field strength of $A_\mu$, $M_{\rm pl}$ is the five-dimensional Planck mass, and $V_g(\lambda)$, $ V_{f0}(\lambda)$, $\kappa(\lambda)$ and $w(\lambda)$ are potentials. 

We seek homogeneous and isotropic black brane solutions to the above equations. We employ the ansatz
\begin{eqnarray}
\textstyle ds^2 &=& e^{2A(r)}\left(-f(r)dt^2 + d\vec x^2+\frac{dr^2}{f(r)}\right)  \\
A_\mu(r) &=& A_0(r) \delta_{\mu 0}\ ,
\end{eqnarray}
with $r$ the radial coordinate of the geometry. The boundary resides at $r = 0$, which is where the field theory lives. 

\begin{figure*}[t]
\begin{center}
\includegraphics[width=0.47\textwidth]{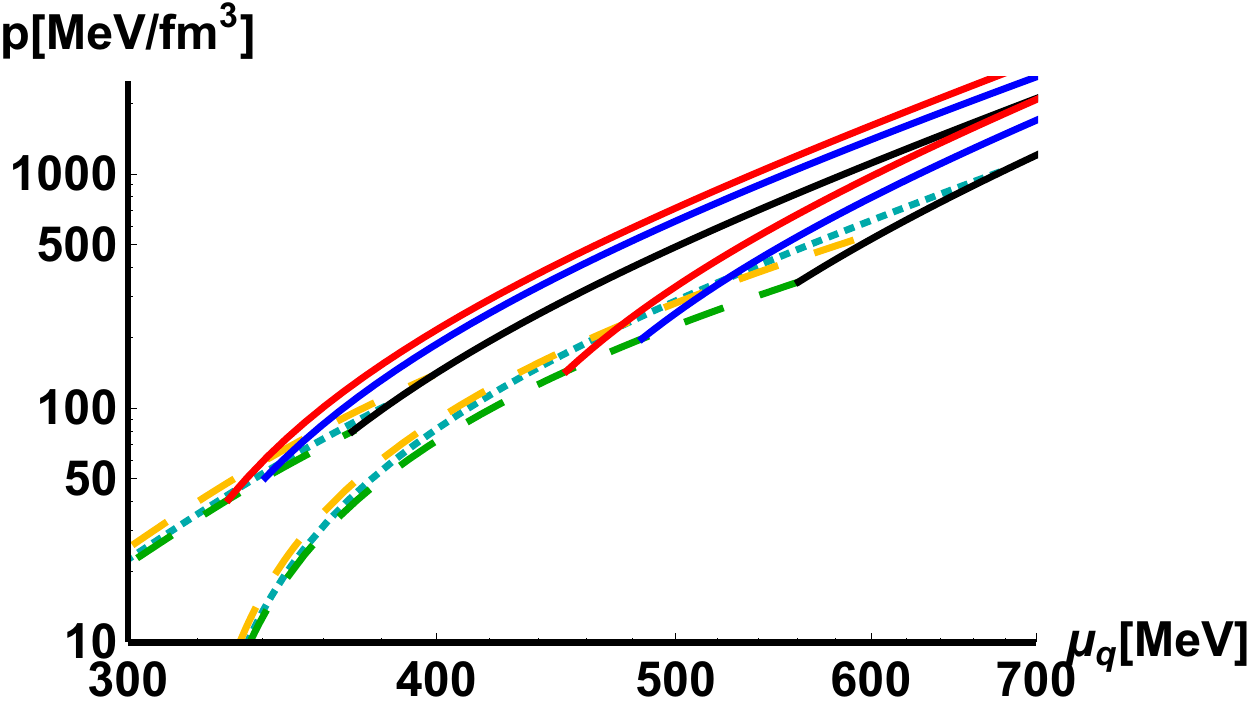}
\includegraphics[width=0.47\textwidth]{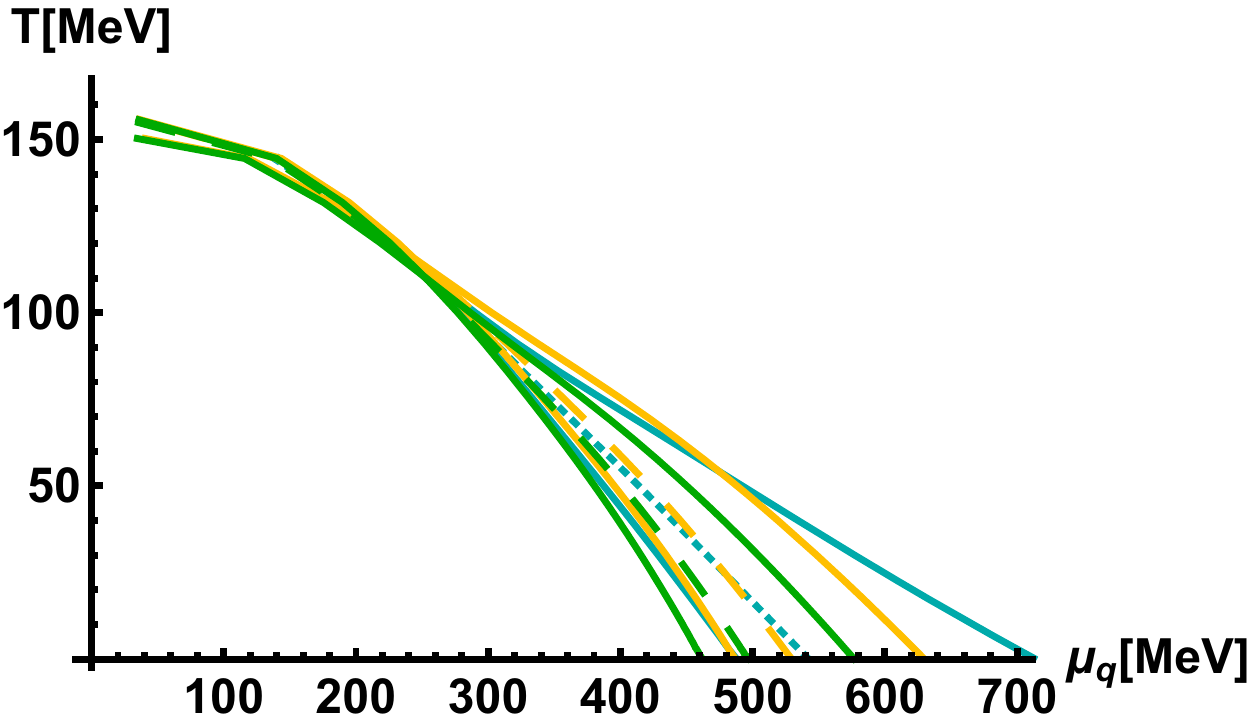}
\caption{Left: The pressures of the DD2 (dashed green), IUF (dotted turquoise) and SFHx (dashed orange) nuclear matter models, shown together with those obtained with the {\bf{5b}} (red solid curve), {\bf{7a}} (blue solid), and {\bf{8b}} (black solid) potentials of V-QCD, for the temperatures of 6.9 (lower) and 75.9 MeV (upper curves). Right: The phase diagram of QCD as suggested by the comparison of the nuclear and quark matter pressures utilized in our work. Corresponding to each of the three nuclear matter EoSs (same color coding as in the left figure), we have three curves corresponding to the V-QCD potentials {\bf{5b}} (left solid curves), {\bf{7a}} (middle dashed or dotted curves), and {\bf{8b}} (right solid curves).} \label{fig:1}
\end{center}
\end{figure*}

Boundary conditions must be imposed at $r = 0$.  For the metric, we impose the boundary condition that the geometry at $r = 0$ is conformally equivalent to (3+1)-dimensional Minkowski spacetime.
The gauge field is dual to the baryon current, and the boundary value of $A_0$ corresponds to the quark chemical potential $\mu_q$,
\begin{equation}
A_0|_{r = 0} = \mu_q \ .
\end{equation}
As such, with $\mu_q \neq 0$ the gauge field is nonzero and the corresponding black brane solution is charged. 
The tachyon $\tau$ is on the other hand dual to the condensate $\bar q q$, and its boundary value sources the bare mass of quarks.
We choose to neglect the bare quark masses of all three lightest quarks, which also means that the beta equilibrium and charge neutrality conditions are automatically attained.  

With vanishing quark masses, the tachyon behaves close to the boundary as \cite{Jarvinen:2015ofa}
\begin{equation}
\tau|_{r\to 0} \simeq \sigma r^3 \ ,
\end{equation}
where $\sigma$ is proportional to the chiral condensate. Interestingly, it turns out that the dominant phase in the setup is the chirally symmetric one with $\tau=0$ throughout the bulk geometry. Finally, the dilaton is dual to the 't Hooft coupling $g^2N_c$ of the Yang-Mills theory. Near $r = 0$, we impose the boundary condition that the dilaton has the expansion
\be
\lambda = -\frac{1}{b_0\log(r\Lambda_{\rm{UV}})}-\frac{8b_1\log[-\log(r\Lambda_{\rm{UV}})]}{9b_0^2\log(r\Lambda_{\rm{UV}})^2}+\ldots\ ,
\ee
where $\Lambda_{\rm UV}$ is an energy scale and $b_0=3,b_1=7/2$ are the coefficients of the QCD beta function in the Veneziano limit: $\beta(g^2 N_c )=-b_0(g^2 N_c)^2 +b_1 (g^2 N_c)^3+\dots$.

For given potentials $V_g(\lambda), V_{f0}(\lambda)$, $\kappa(\lambda)$, $w(\lambda)$ and boundary conditions, we may next proceed to solve the equations of motion for charged black brane solutions. The Hawking temperature $T$ of the black brane corresponds to the temperature of the dual quark matter. For given $\mu_q$ and $T$, the quark matter pressure $p(\mu_q,T)$ in the grand canonical ensemble can then be obtained by evaluating the on-shell action (\ref{eq:SVQCD}), together with appropriate counterterms and the Gibbons-Hawking boundary action \cite{Gursoy:2007er,Jarvinen:2015ofa}. We note that the Plank mass $M_{\rm pl}$ and the energy scale $\Lambda_{\rm UV}$ determine the overall normalization 
of the pressure.

\begin{figure*}[t]
\begin{center}
\includegraphics[width=0.43\textwidth]{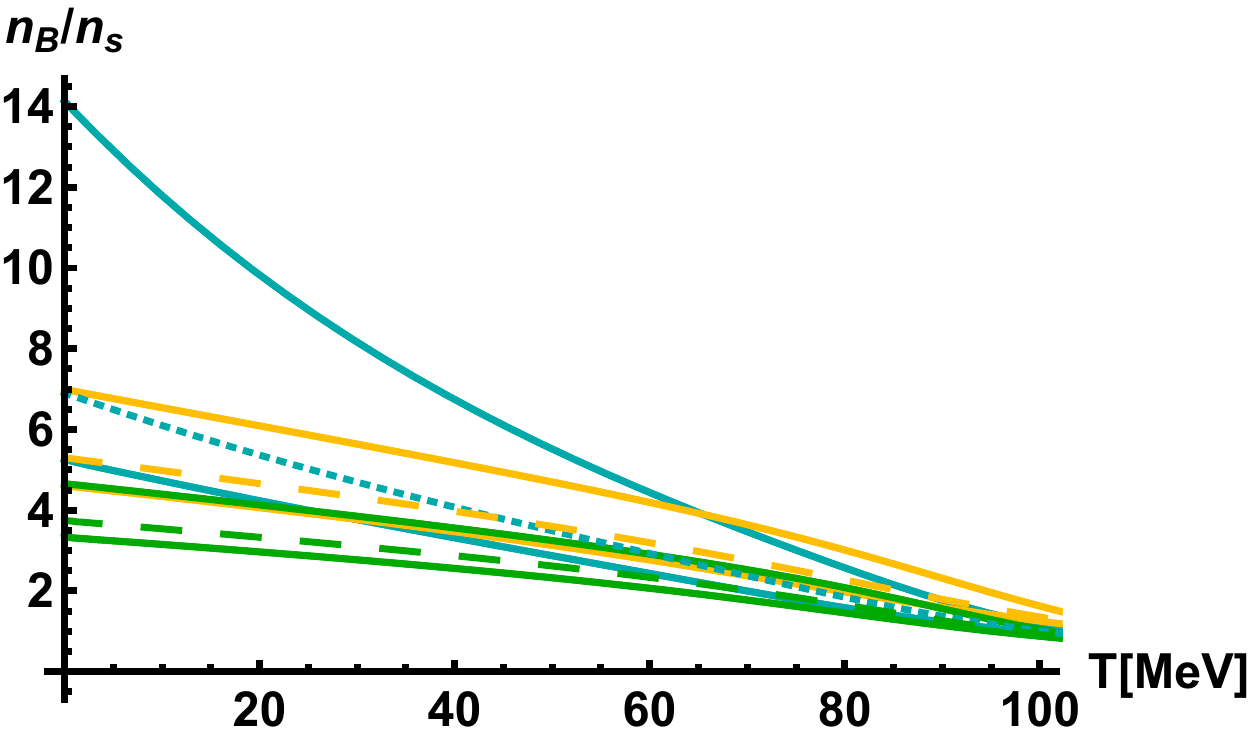}
\includegraphics[width=0.47\textwidth]{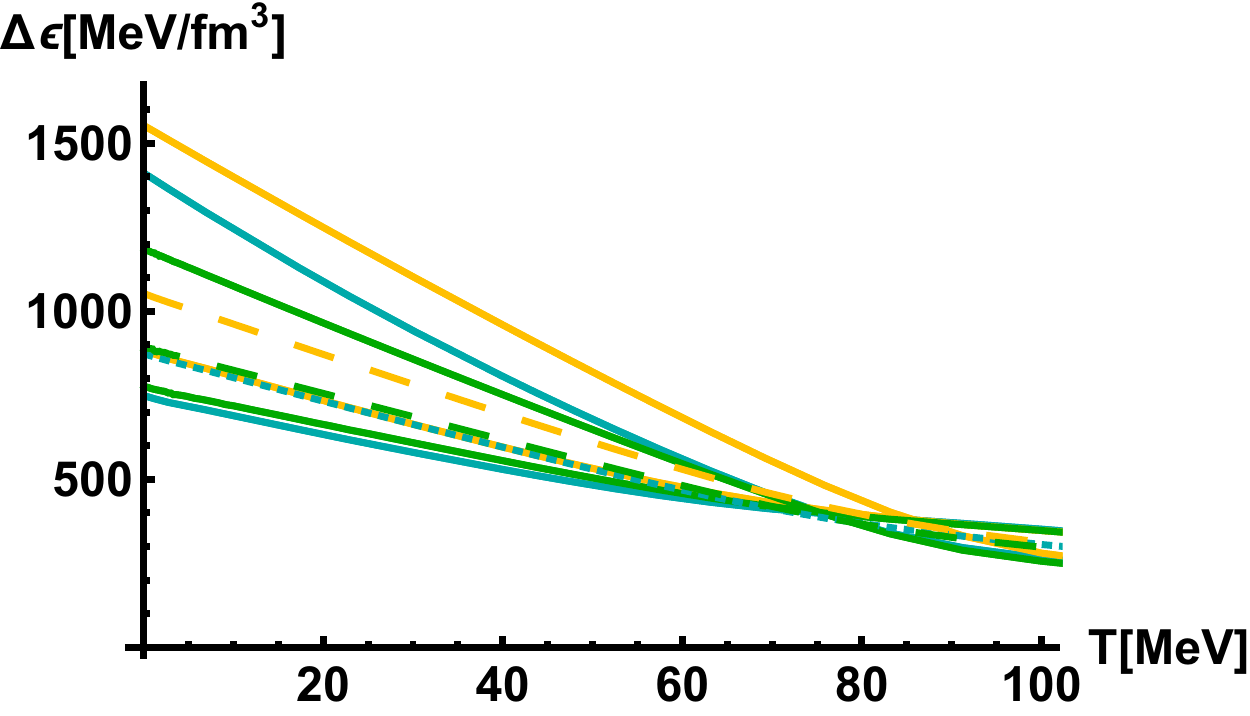}
\caption{Left: The transition density as a function of temperature, given for the three nuclear physics models considered (same color coding as before). For each nuclear EoS, we again have three curves corresponding to the V-QCD potentials {\bf{5b}} (lower solid curves),  {\bf{7a}} (middle dashed or dotted curves), and {\bf{8b}} (upper solid curves). Right: The latent heat of the (first order) deconfinement transition as a function of temperature. The notation  follows the left figure.} \label{fig:2}
\end{center}
\end{figure*}

The potentials $V_g(\lambda), V_{f0}(\lambda)$, $\kappa(\lambda)$ and $w(\lambda)$ are 
constrained by matching onto QCD thermodynamics at vanishing density. Their functional forms read
\bea
\nonumber 
\frac{V_g}{12}& =& \textstyle  1+V_1\lambda+\frac{V_2\lambda^2}{1+\frac{\lambda}{\lambda_0}}+V_{\rm{IR}}e^{-\frac{\lambda_0}{\lambda}}\left(\frac{\lambda}{\lambda_0}\right)^{\frac{4}{3}}\sqrt{1+\frac{\lambda}{\lambda_0}} \ , \\
V_{f0} & = & \textstyle W_0+W_1\lambda+\frac{W_2\lambda^2}{1+\frac{\lambda}{\lambda_0}}+W_{\rm{IR}}e^{-\frac{\lambda_0}{\lambda}}\left(\frac{\lambda}{\lambda_0}\right)^2, \nonumber\\
\textstyle  \frac{w_0^{-1}}{w(\lambda)} & = &\textstyle 1+\frac{w_1 \frac{\lambda}{\lambda_0}}{1+\frac{\lambda}{\lambda_0}}+\bar w_0e^{-\frac{\lambda_0}{w_s\lambda}}\frac{\left(\frac{w_s\lambda}{\lambda_0}\right)^{\frac{4}{3}}}{\log(1+\frac{w_s\lambda}{\lambda_0})} \label{eq:potentials}\\ \nonumber
\textstyle  \frac{\kappa_0^{-1}}{\kappa(\lambda)} & = & \textstyle 1+\kappa_1\lambda+\bar\kappa_0\left(1+\frac{\bar\kappa_1\lambda_0}{\lambda}\right)e^{-\frac{\lambda_0}{\lambda}}\frac{\left(\frac{\lambda}{\lambda_0}\right)^{\frac{4}{3}}}{\sqrt{\log(1+\frac{\lambda}{\lambda_0})}} \nonumber \ ,
\eea
where $V_1$, $V_2$, $V_{\rm IR}$, $\lambda_0$, $W_0$, $W_1$, $W_2$, $\kappa_0$, $W_{\rm{IR}}$, $w_0$, $w_1$, $\bar w_0$, $w_s$, $\bar\kappa_0$, and $\bar\kappa_1$
are parameters.  
As shown in 
\cite{Gursoy:2009jd}, the parameters $V_1$, $V_2$, $V_{\rm IR}$ and $\lambda_0$ in the potential $V_g(\lambda)$ can be determined by matching to pure Yang-Mills theory both on the lattice and at weak coupling \cite{Panero:2009tv}. 
In the presence of quarks,  lattice data at vanishing chemical potentials constrain $V_{f0}(\lambda)$ by the interaction measure \cite{Borsanyi:2013bia} and $w(\lambda)$ by the baryon number susceptibility \cite{Borsanyi:2011sw}. 
In particular, the UV dimension of the $ q\bar q$ operator and the RG flow of the quark mass and the coupling fix $W_1$ ,$W_2$, $\kappa_0$, and $\kappa_1$. 
The potential for the tachyon kinetic term, $\kappa(\lambda)$, is insensitive to fitting to lattice QCD data. However, in the dominant phase probed in this work $\tau=0$, so also the thermodynamic quantities are insensitive to the parameter values in $\kappa(\lambda)$.  We are therefore left with $W_0, W_{\rm{IR}}$, $w_0$, $w_1$, $\bar w_0$, $w_s$, $\bar\kappa_0$, and $\bar\kappa_1$ as the free parameters of the potentials.

In Ref.~\cite{Jokela:2018ers}, it was demonstrated that at $\mu_q > 0$ and $T = 0$, our holographic model automatically  leads to thermodynamic results which fall in a very reasonable range. Out of the potentials studied in \cite{Jokela:2018ers}, we focus on those that are not in conflict with any robust NS observations, implying in particular that they do not predict a strong first order deconfinement transition at such a low density that two-solar-mass stars would not exist. This singles out the potentials in (\ref{eq:potentials}) for which $W_0\ne 0$, denoted by {\bf{4}}--{\bf{9}} in \cite{Jokela:2018ers}, of which we choose as three representative examples the potentials {\bf{5b}} and {\bf{7a}}, and {\bf{8b}}.  The parameters corresponding to these three potential choices can be found in Appendix~A.2 of \cite{Jokela:2018ers}.
Of these, {\bf{5b}} and {\bf{8b}} correspond to the maximum allowed variance, while the potential {\bf{7a}} can be considered a typical, or average, V-QCD prediction. We note that the potential {\bf{7a}} also leads to phenomenologically reasonable baryon physics, as recently discussed in \cite{Ishii:2019gta}.

\section{Results}\label{sec:results}

Having established the procedure, with which we describe the confined and deconfined phases of QCD, let us proceed to inspect the resulting thermodynamic properties of NS matter at different temperatures. Fig.~\ref{fig:1} (left) demonstrates our basic procedure for sorting out the phase structure of the theory. For two different temperatures, chosen as 6.9 and 73 MeV for illustrative purposes, we compare here the grand canonical pressures of the DD2, IUF, and SFHx nuclear matter models with those obtained using the V-QCD potentials {\bf{5b}}, {\bf{7a}}, and {\bf{8b}} of  \cite{Jokela:2018ers} --- all evaluated in beta equilibrium as explained above. In each case, we assume the phase transition from the nuclear to the quark matter phase to occur at the quark chemical potential, where the two curves meet. This implies that we ignore the so-called mixed-phases scenario that would become relevant if the (unknown) microscopic surface tension of high-density QCD, describing a domain wall separating its confined and deconfined phases, was sufficiently low. In the present work, we make this choice for the sake of simplicity, but note that it would be useful to address the so-called Gibbs construction featuring mixed phases in future work.

Repeating the above procedure for tens of different temperatures, we obtain a set of points to mark the phase transition line on the phase diagram of the theory, shown in Fig.~\ref{fig:1} (right). For each nuclear matter EoS, we display three curves corresponding to the three V-QCD potentials, such that the left- and right-most curves stand for the potentials {\bf{5b}} and {\bf{8b}}, and the middle curve to {\bf{7a}}. In a loose sense, these ``bands'' can be considered uncertainty estimates for our results, given a fixed low-energy EoS. As is evident from this figure, at temperatures $T \gtrsim 75$ MeV all combinations of nuclear matter EoSs and V-QCD potentials yield nearly identical phase diagrams, which all end at a temperature around $T=155$ MeV, beyond which a phase transition no longer exists. However, in the zero-temperature limit the transition chemical potential varies between 460 and 700 MeV. We note that the main source of this variation originates from the V-QCD potentials and not the nuclear EoSs.\footnote{This variation can be traced to the normalization of the bulk gauge field, $w(\lambda)$. Fixing it would require the existence of robust lattice data either at nonzero $\mu_q$ or external magnetic field $\mathbf{B}$.}

In Fig.~\ref{fig:2}, we next plot the transition baryon number density  $n_B$ (left panel) and latent heat $\Delta \epsilon$ (right panel), in both of which the lowest curves of each color correspond to the potential {\bf{5b}} and the uppermost ones to {\bf{8b}}. Depending on the nuclear matter model and V-QCD potential, the $T = 0$ transition baryon number density lies between 3 and 14$n_s$, with $n_s$ the nuclear saturation density. Here, it turns out that the IUF model is responsible for the largest values. In contrast, the latent heat at $T = 0$ varies only by about a factor of two, lying between 700 and 1500 MeV/fm$^3$.  The latent heat is thus of the order of the energy density of nuclear matter, which indicates a strong first order phase transition.  For all nuclear models and V-QCD potentials, both the transition baryon number density and latent heat decrease as $T$ increases.  Likewise, the variance of both the transition baryon number density and latent heat decrease as $T$ increases. 

\begin{figure}[t]
\begin{center}
\includegraphics[width=0.43\textwidth]{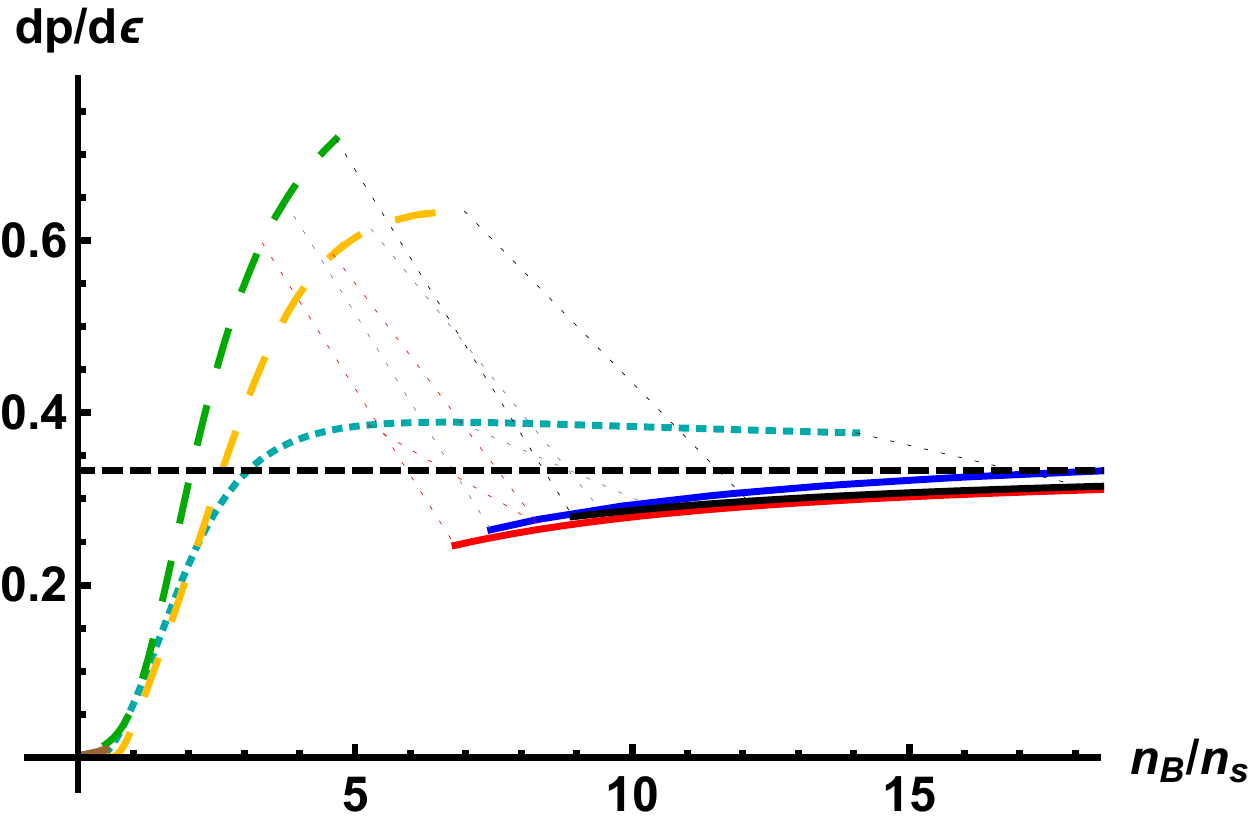}
\caption{The speed of sound squared $c_s^2$, or stiffness $dp/d\epsilon$, as a function of baryon density for the nuclear matter models and V-QCD potentials at vanishing temperature. The color codings follow the choices made in Fig.~\ref{fig:1} (left), while the dotted lines indicate the transitions between the two descriptions in all combinations of models and potentials (see also Table \ref{table:transition}).} \label{fig:4}
\end{center}
\end{figure}

\begin{figure}[t]
\begin{center}
\includegraphics[width=0.43\textwidth]{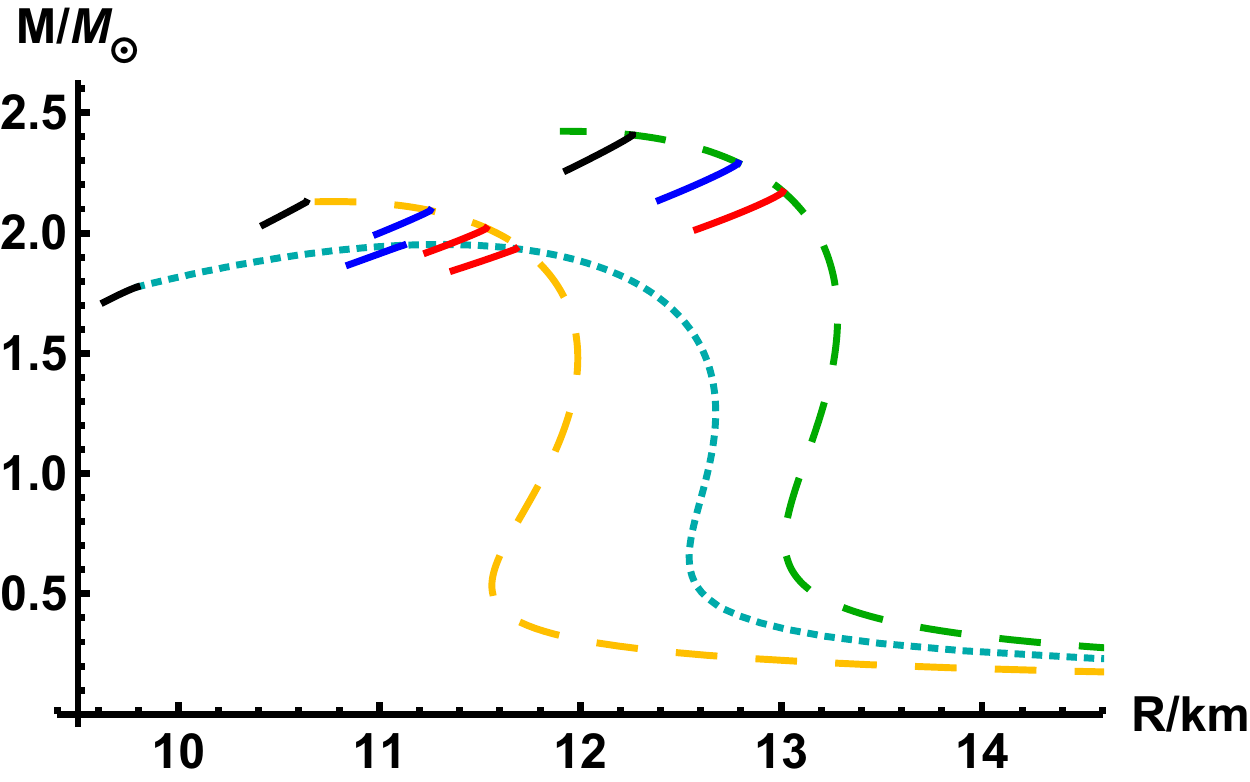}
\caption{The mass-radius curves resulting from each of the three nuclear matter models considered, with the color scheme following the previous figures. The straight line segments correspond to the unstable branches, where the cores are populated by holographic quark matter described by the different V-QCD potentials.} \label{fig:3}
\end{center}
\end{figure}

Returning momentarily to the limit of zero temperature, we display the speed of sound squared $c_s^2=dp/d\epsilon$ and MR-relation originating from each of the nuclear matter models and V-QCD potentials in Figs.~\ref{fig:4} and \ref{fig:3}, respectively (see also Table \ref{table:transition}, relevant for the speed of sound plot). In the latter, the most interesting quantity is clearly the corresponding maximal mass of stable NSs, where the MR-curves either bend down or end due to a strong first order phase transition.  To date, the heaviest commonly accepted NS mass measurement reads $M = 2.01 \pm 0.04 M_\odot$ \cite{Antoniadis:2013pzd}, but one should in addition recall the very recent detection of an extraordinarily massive NS with $M=2.17 \pm 0.1 M_\odot$ \cite{Cromartie:2019kug}. Combining the latter measurement with claims of the EM counterpart of GW170817 constraining the maximum NS mass from above by $2.16\pm 0.16 M_\odot$ \cite{Margalit:2017dij,Rezzolla:2017aly}, we are left with the conclusion that the maximal mass of stable NSs should fall within the range 2.07--2.33$M_\odot$.

Due to the large latent heats obtained in our setup, the presence of a quark matter core in a quiescent ($T=0$) NS always results in the star becoming unstable to gravitational collapse.  
To this end, the NS mass, for which a quark matter core just begins to form, i.e.~where the central density of the NS reaches the critical one, uniquely determines the maximum mass with the exception of one case (model IUF with potential {\bf 8b}), where the star becomes unstable before the onset of quark matter.  In Table~\ref{table:masses}, we display the maximal masses for all combinations of the three nuclear matter EoSs and V-QCD potentials, as well as for the pure nuclear matter cases. From here, we see that the potential {\bf 8b} has a negligible effect on the maximum NS mass for all nuclear EoSs, which is a direct consequence of the fact that the transition density is always relatively large for this potential, cf.~Fig.~\ref{fig:2} (left).  In contrast, the potentials {\bf 5b} and {\bf 7a} reduce the maximum neutron star mass by as much as 15\%, although for the IUF model this effect is smaller, again due to large transition densities.

\begin{table*}
	\begin{tabularx}{0.8\textwidth}{cYYY} \toprule
		$p$, $\epsilon$ [MeV/fm$^3$], $n$[$n_s$] & Potential {\bf{5b}} & Potential {\bf{7a}} & Potential {\bf{8b}} \\
		\midrule
		DD2  & 159, 578, 3.33 & 217, 671, 3.73 & 380, 908, 4.65  \\
		IUF  & 253, 973, 5.24 & 411, 1379, 6.88 & 1246, 3561, 14.06  \\
		SFHx	& 250, 825, 4.60 & 351, 993, 5.29 & 645, 1462, 6.98  \\
		\bottomrule
	\end{tabularx}
	\caption{The values of the pressure, energy density, and baryon density at the $T=0$ transition from nuclear to quark matter, given separately for all combinations of nuclear matter models and V-QCD potentials.}\label{table:transition}
\end{table*}

\begin{table}
	\begin{tabularx}{0.48\textwidth}{cYYYY} \toprule
		$M_\text{max}/M_\odot$ & Pot.~{\bf{5b}} & Pot.~{\bf{7a}} & Pot.~{\bf{8b}} & Pure NM \\
		\midrule
		DD2  & 2.17 & 2.29 & 2.41 & 2.42  \\
		IUF  & 1.93 & 1.95 & 1.95 & 1.95 \\
		SFHx	& 2.02 & 2.09 & 2.13 & 2.13  \\
		\bottomrule
	\end{tabularx}
	\caption{Maximal masses of NSs (at $T=0$) built using different nuclear matter models and V-QCD potentials for the quark matter phase. The ``Pure NM'' column refers to NSs built purely from the low-density EoS, i.e.~with no transition to quark matter phase at all. Note that the maximal masses in the other cases originate from the star becoming unstable upon the central density reaching the critical density for the deconfinement transition in our ``Maxwell construction'' setup.}\label{table:masses}
\end{table}

Strictly enforcing the existence of the $M = 2.01 \pm 0.04 M_\odot$ NS (barely) rules out the IUF model both with and without any of the three V-QCD potentials. With these EoSs excluded, the variance of the phase diagram in Fig.~\ref{fig:1} greatly decreases, with the  $T = 0$ transition chemical potentials now lying in the range 460--630 MeV and the corresponding baryon number densities in the range 2.5--7$n_s$. Taking the constraint $M_\text{max}/M_\odot\in [2.07,2.33]$ further into account, we also remove the combinations DD2 and {\bf 8b} as well as SFHx and {\bf 5b}. After this,  the transition density is constrained to lie between 3.3 and 7.0$n_s$ and the latent heat between 770 and 1550 MeV/fm$^3$. To assess the universality of these findings, it would clearly be very interesting to study how they get modified with other holographic models of quark matter, such as the Sakai-Sugimoto model \cite{BitaghsirFadafan:2018uzs}.

Finally, we note in passing that we have also studied the behavior of the so-called thermal index $\Gamma_\text{th}\equiv 1+\frac{\Delta p}{\Delta \epsilon}$, where $\Delta x \equiv x-x|_{T=0}$ for fixed baryon density, with all three nuclear matter models considered. This quantity was recently determined for the first time in an ab initio calculation in \cite{Carbone:2019pkr} (see also the related work \cite{Constantinou:2015mna}), giving us an opportunity to perform a valuable cross-check of the low-density behaviors of our model EoSs. The result of this exercise was encouraging, as we witnessed all three EoSs reproducing the qualitative behavior of the findings of  \cite{Carbone:2019pkr} at both $T=20$ and 50 MeV.

\section{Discussion}\label{sec:discussion}

We have approached the problem of building realistic finite-$T$ EoSs for neutron star merger simulations using different state-of-the-art descriptions of the confined and deconfined phases of QCD. On the low-density side, we have employed three well-known model setups available on the market -- DD2 \cite{Typel:2009sy}, IUF \cite{Fattoyev:2010mx}, and  SFHx \cite{Steiner:2012rk} -- which are all at least marginally compatible with robust observational information on NS properties and for which tabulated EoSs are available for a wide range of temperatures. On the high-density side, we have on the other hand used different phenomenological potentials within the V-QCD model of \cite{Gursoy:2007cb,Gursoy:2009jd,Jarvinen:2011qe,Alho:2013hsa,Jokela:2018ers}, which is a highly developed holographic framework for the description of quark matter, built by fitting the associated potentials to lattice QCD data at zero or small density. 

On both the nuclear and the quark matter sides, the models and V-QCD potentials were chosen to produce maximal allowed variance for the thermodynamic properties of QCD, which resulted in sizable differences for quantities such as the $T=0$ transition density and the size of the latent heat. On the contrary, the qualitative form of the phase diagram -- and even the quantitative location of the transition line at higher temperatures -- was seen to be highly model-independent. A more detailed analysis of the maximal masses of stable NS solutions allowed us to further discriminate between the EoSs, leaving only four combinations of nuclear matter models and V-QCD potentials intact.

The fact that NS observations constrain the possible behavior of the $T=0$ NS matter EoS significantly is not a new discovery (see e.g.~\cite{Annala:2017llu,Most:2018hfd,Annala:2019puf} for related discussions), but the fact that this observation pertains to nonzero temperatures is a very interesting and nontrivial result. The four EoS combinations we have singled out above represent very different, yet observationally viable, behaviors of NS matter and are immediately amenable to use in simulations. The viable low-density EoSs, i.e~DD2 and SFHx, are tabulated for several fixed electron fractions $Y_e$ e.g.~in \cite{Hempelpage}, while the quark matter EoSs we have constructed in this work are similarly tabulated in the ancillary material of this paper.

\medskip

\begin{acknowledgments}
We thank Arianna Carbone, Matthias Hempel, Matti J\"arvinen, Jere Remes, and Larry Yaffe for useful discussions. The research of N.~J.~and A.~V.~has been supported by the European Research Council, grant no.~725369, and by the Academy of Finland grants no. 1322307 and 1322507.  P.~C.~and A.~L.~are supported by the Black Hole Initiative at Harvard University, which is funded by a grant from the John Templeton Foundation.
\end{acknowledgments}

\medskip

\bibliographystyle{apsrev4-1}

\bibliography{biblio}

\end{document}